\documentstyle[prd,aps]{revtex}
\begin{document}
\title{Effect of spatial fluctuations on the scaling of the scaled factorial
moments in second-order quark-hadron phase transition within Ginzburg-Landau
description}
\author{C.B. Yang$^{1,2}$ and X. Cai$^{1,3}$} 
\address{$^1$ Institute of 
Particle Physics, Hua-Zhong Normal University, 
Wuhan 430079, the People's Republic of China\\
$^2$ Theory Division, RMKI, KFKI,
 Budapest 114., Pf. 49, H-1525 Hungary\\
$^3$ Physics Department, Hubei University, Wuhan 430062, People's
Republic of China}
\date{\today}
\maketitle

\begin{abstract}
The scaled factorial moments in second-order quark-hadron
phase transition are reexamined within the Ginzburg-Landau description,
with the spatial fluctuations of phase angle of the complex field $\phi$
taken into account rigorously. Scaling behaviors between $F_q$ and
$F_2$ are shown, and the exponent $\nu$ is found very close to the one
without spatial fluctuations. 

{{\bf PACS} number(s): 13.85.Hd, 05.70.Fh, 12.38.Mh}
\end{abstract}

The idea of scaled factorial moments [1] has been used since being
proposed to study dynamical fluctuations extensively in both
theoretical investigations and experimental analyses due to the powerful
ability to eliminate statistical (non-dynamical) fluctuations. Some
scaling laws for the moments have been predicted theoretically and
observed in high energy hadronic and nuclear interactions [2]. In high
energy heavy-ion collisions, a new matter state --quark-gluon plasma
(QGP)-- may be produced. The newly produced hot system will cool with
its expanding and will undergo a phase transition from the deconfined QGP
to confined hadrons which will be detected in experiments. Since the
existence of the phase transition is associated with properties of
the nontrivial chromodynamical vacuum, the study of quark-hadron phase
transition has been a hot point in both particle physics and nuclear
physics for more than a decade.
Besides the unique features of quantum chromodynamics, the lack
of control of the temperature in the phase transition distinguishes
the problem from the standard critical phenomena such as ferromagnetism.
The nonperturbative nature of hadronization process in the phase transition
precludes at this stage any observable hadronic predictions from first
principles. Hence, among many phenomenological models, the Ginzburg-Landau
description has been used as a framework to make predictions, which can be
compared with experiment [3, 4], on the hadronic observables. Up to now,
this model has been used in the study of various scaling behaviors in
the phase transition, such as those of the scaled factorial moments in
both first-order [5-7] and second-order [8,9] phase transitions, the
multiplicity difference correlators [10-12], the multiplicity distributions
[13,14], etc.

In the Ginzburg-Landau description of second-order phase transition, 
the scaled factorial moments can be expressed as [9]
\begin{equation}
F_q=f_q/f_1^q, \mbox{\hspace{0.8cm}} f_q={1\over Z}\int {\cal D}\phi
\left(\int dz\mid\phi\mid^2\right)^q \exp(-F[\phi])
\end{equation}

\noindent with $Z=\int {\cal D}\phi \exp(-F[\phi])$, the free energy functional
$F[\phi]=\int dz [a\mid\phi\mid^2+b\mid\phi\mid^4+c\mid\bigtriangledown
\phi\mid^2]$, $a\propto (T-T_C)$ representing the distance from the critical
point, $b$ and $c$ larger than zero. Here $|\phi|^2$ is associated with
the multiplicity density of the system. Similar expressions can be derived
for other quantities mentioned above. In all former studies of second-order
phase transition the gradient term is simply taken to be zero, i.e., the
field $\phi$ is regarded spatially uniform. The spatial integral of the
functional over a two-dimensional bin with size $\delta^2$ is then
$F[\phi]=\delta^2(a\mid\phi^2\mid+b\mid\phi\mid^4)$ (Calculations based on 
this functional will be referred to mode 1 in this paper). This is of course a
very crude approximation. The advantage of such an approximation is,
however, that it turns the functional integration into a normal one. Thus,
the calculation becomes quite easy under the approximation. Numerical
results do not show the so-called intermittency behavior, but the 
$F$-scaling, $F_q\propto F_2^{\beta_q}$ with universal scaling law $\beta_q
=(q-1)^\nu$, is shown to be valid. The exponent $\nu$ is called as a
universal one in the sense that it is insensitive to the values of the
parameters in the model and that it is completely determined by the structure
of the functional concerned.

The contribution from the gradient term to the moments and the exponent $\nu$ 
should be evaluated in some way. Once the gradient term is taken into the
functional, one is faced with serious difficulty in the calculations,
considering the fact that the value of parameter $b$ for the $\phi^4$ term
can be determined in no way from first principles or from experimental input
and may be very large. Even the parameter $b$ is indeed very small, negative
value of $a$ in our interested region also excludes the possibility of
performing usual perturbative calculations. The role played by the gradient
term is investigated in Refs. [15, 16]. In Ref. [15] $\phi$ in each bin is
still uniform, but the values of $\phi$ in all neighboring bins are taken to
be $\phi_0$, field configuration corresponding to the minimum of ``potential''
$V(\phi)\equiv a\mid\phi\mid^2+b\mid\phi\mid^4$.
So the square of the gradient of $\phi$ is $\delta^{-2}(\phi-\phi_0)^2$. This
approximation also transforms the functional integration into a normal one. 
Numerical results show that the universal scaling law $\beta_q=(q-1)^\nu$ is 
still valid and that the exponent $\nu$ is almost the same as without the
gradient term. In Ref. [16] the details of spatial fluctuations of $\phi$ in
a bin is simulated by the Ising model for one-component spins $s$. Each bin is
assumed large enough to contain several spin sites. The multiplicity
in a bin is associated with the sum of spins on all sites in the bin by
\begin{displaymath}
n_i=\lambda \mid\sum_{j\in {\rm bin}}s_j\mid^2 \theta(\sum_{j\in {\rm bin}}
s_j)\ ,
\end{displaymath}

\noindent in which scale factor $\lambda$ relates the hadron density to
the lattice spin density in the Ising model. Since $\lambda$ relates a physical
space to a mathematical space  and there is no way to determine it a priori,
several $\lambda$'s are used in Ref. [16]. This time, the exponent $\nu$
depends on the unknown temperature, and, after averaging over the temperature,
$\nu$ is still in the range given in Refs. [8, 9, 15].

Though the simulation in Ref. [16] is convincing, it is for lattice with 
one-component spins. In the Ginzburg-Landau model for second-order phase 
transition, the field $\phi$ is a complex number, or in other words,
$\phi$ has two components. Thus at first glimps the simulation in Ref. [16] 
does not correspond to the real problem discussed in the Ginzburg-Landau 
model. But as will be shown soon the simulation relates to the physics
in an indirect way. The simulation, however, is still useful because
one can translate the two-component field into a one-component
one, as will be illustrated below in this paper. 

In this paper, it is tempted to investigate the universality of the exponent
$\nu$, with the spatial fluctuations of the phase angle of the complex
field $\phi$ fully taken into account. The magnitude fluctuations of $\phi$
is neglected in this investigation for reasons which will be given below.
As will be seen soon, the contribution from spatial fluctuations of the phase
angle of the field $\phi$ can be taken into account in a complete and 
rigorous way, and the integration over the spatial fluctuations of the phase
angle of the field $\phi$ will reduce the problem to one with one-component
field. 

Our first observation is that all terms except the gradient term in the
functional integral of Eq. (1) depend only on $\mid\phi\mid^2$. Then it is
convenient to write the two-component field $\phi$ as a complex number in the
form $\phi= \phi_{\rm R}\exp(i\phi_{\rm I})$. The spatial fluctuations of 
the field can
be those of the magnitude $\phi_{\rm R}$ and/or of the phase angle $\phi_{\rm
I}$ (or orientation in an abstract space). The gradient term turns out to be
\begin{equation}
\mid\bigtriangledown\phi\mid^2=(\bigtriangledown \phi_{\rm R})^2
+\phi_{\rm R}^2(\bigtriangledown\phi_{\rm I})^2\ .
\end{equation}.

\noindent If $\bigtriangledown\phi_{\rm I}$ is fixed in the process (either
because of the uniformity of $\phi_{\rm I}$ or uniform change of $\phi_{\rm
I}$), the effect of $(\bigtriangledown\phi_{\rm I})^2$ can be effectively
considered by a new parameter $a^\prime$ in place of $a$ with $a^\prime=a
+c(\bigtriangledown\phi_{\rm I})^2$, reducing the problem to former case if
the $\bigtriangledown \phi_{\rm R}$ term is also neglected. Needless to
say, such a uniformity can be realized only in the case with very strong
correlation between particles within the bin and, most probably, does not
correspond to our case under study. Generally, the phase angle $\phi_{\rm
I}$ can be in any form, and the full contribution due to its fluctuations
must be evaluated. Fortunately, the integral over $\phi_{\rm I}$ can
be carried out easily since it is of Gaussian form. Using a formulus
\begin{displaymath}
\int \phi_{\rm R}{\cal D}\phi_{\rm R}{\cal D}\phi_{\rm I}
\exp[-\int dx \phi_{\rm R}^2(\bigtriangledown \phi_{\rm I})^2]={\rm const}
\cdot \int {\cal D}\phi_{\rm R}\ ,
\end{displaymath}
\noindent one transforms the two-fold functional integral into a one-fold
one. The unimportant constant will be cancelled in the expression for $f_q$.
Then one has
\begin{equation}
f_q={\int {\cal D}\phi_{\rm R}
\left(\int_\delta dz\phi_{\rm R}^2\right)^q \exp(-F[\phi_{\rm R}])\over
\int {\cal D}\phi_{\rm R}\exp(-F[\phi_{\rm R}])}\ ,
\end{equation}

\noindent with functional $F[\phi_{\rm R}]$ exactly the same form as the
original $F[\phi]$. The important difference between this expression from
Eq. (1) is that the functional integral variable in this new expression is
a real function instead of a complex function in Eq. (1). Similar expression
can also be derived for the case of first-order phase transition, with an
extended functional $F[\phi_{\rm R}]$.

Now we take the field $\phi_{\rm R}$ (magnitude of $\phi$) uniform, or in
other words, the gradient term of $\phi_{\rm R}$ is omitted. (Calculations
based on this approximation will be referred to mode 2 in this paper.) Based
on the work Ref. [16] one can drop off the $\bigtriangledown \phi_{\rm R}$
term, because the problem now is exactly within the scope of Ref. [16], and
the conclusions in Ref. [16] encourage us to neglect the spatial fluctuations
of $\phi_{\rm R}$ as long as the universal scaling exponent $\nu$ is concerned. Then one gets the factorial moments as functions of variable $x$
\begin{equation}
f_q={\int_0^{\infty} dy y^{2q}\exp(xy^2-y^4)\over \int_0^{\infty} dy
\exp(xy^2-y^4)}\ ,
\end{equation}

\noindent with $x=a\delta^{3/2}/b^{1/4}$. From this expression the scaled
factorial moments $\ln F_q$ can be calculated, and the results are shown as
functions of $-\ln x$ in Fig. 1 for $q$ from 2 to 8 within the range 
$x\in (0.5, 4.0)$. One can see clearly that no strict intermittency can be 
claimed since all $F_q$ approach finite values in the limit $x\to 0$.
So, no intermittency is shown in the phase transition, as shown in
former studies. More importantly, the power law can be found between $F_q$
and $F_2$, as shown in Fig. 2 with the same data as in Fig. 1. 

For the convenience of comparison with former case, we write down the
expressions of the scaled factorial moments without spatial fluctuations
(mode 1 in this paper), which can be read
\begin{equation}
f_q={\int_0^\infty dy y^q
\exp(xy-y^2)\over \int_0^\infty dy \exp(xy-y^2)}\ ,
\end{equation}

\noindent with $x=a\delta/\sqrt{b}$. Numerical results for $\ln F_q$ in
this mode are shown in Fig. 3. In the upper part of the figure $\ln F_q$
are shown as functions of $-\ln x$ for $q$ from 2 to 8 with $x$ in the
same interval $x\in (0.5, 4.0)$, and in the lower part $\ln F_q$ are shown
as functions of $\ln F_2$ with the same data as in upper part. One can see
from upper part of the figure that the general behaviors of $\ln F_q$ as
functions of $-\ln x$ is similar to those in Fig. 1, though the definition
of $x$ in this case is different from that for Fig. 1. The values of $\ln
F_q$ in the two cases are also different. For same value of $x$, $\ln F_q$
in the former case have larger values. This difference is reasonable if
one notices the difference in the definition of variable $x$. What
interests us is the scaling law between $F_q$ and $F_2$. The power law
scaling between $F_q$ and $F_2$ can be seen obviously in the lower part
of Fig. 3, the same as shown in other studies cited in the references. 

From Fig. 2 and the lower part of Fig. 3, one can get the scaling exponents
$\beta_q$ for the two different modes by fitting the curves. $\beta_q$ can
also be given analytically. One can expand the expressions for $\ln F_q$ in
the two modes as power series of $x$ in small $x$ limit, and then one gets
the slopes $K_q$ for $\ln F_q$ and $\beta_q=K_q/K_2$. If the scaling is
good enough as shown in this paper, the exponents obtained from these two
different ways can be regarded equal. The expressions for
$K_q$ for the two modes in this paper are
\begin{eqnarray*}  
&&K_q={\Gamma(q/2+1)\over \Gamma(q/2+1/2)}-q{\Gamma(3/2)\over \Gamma(1)}+(q-1)
{\Gamma(1)\over \Gamma(1/2)}\ \ \ \mbox{for mode 1}\ ,\\
&&K_q={\Gamma(q/2+3/4)\over \Gamma(q/2+1/4)}-q{\Gamma(5/4)\over
\Gamma(3/4)}+(q-1) {\Gamma(3/4)\over \Gamma(1/4)}\ \ \ \mbox{for mode 2}\ .
\end{eqnarray*}

\noindent One can find only a small difference between the exponents
$\nu$ from these two expressions. The results are shown in Fig. 4. In mode 1
(without spatial fluctuations) $\nu$=1.3335, and in mode 2 (with spatial
fluctuations of the phase angle of the field $\phi$) $\nu$=1.2772. The
exponents obtained from these analytical expressions are very close to
the ones from the fitting. The universal exponent $\nu$ are also very
close to one another and can be regarded
as the same within accuracy $4\%$. Physically, these two modes correspond
to different situations. In mode 1 no spatial fluctuation of $\phi$ is in
the problem, but in mode 2 the spatial fluctuations of the phase angle of
the complex field $\phi$ are fully evaluated. Since these two different
considerations give very close exponents $\nu$, one can
say that the exponent $\nu$ is indeed insensitive to the spatial fluctuations.
It should be pointed out that the slight difference of the exponent $\nu$
for mode 1 in this paper from that in Ref. [9] comes from the different $x$
regions from which $\beta_q$ is calculated. In Ref. [9] $\beta_q$ (thus $\nu$)
is obtained by fitting in larger $x$, but in this paper $\beta_q$ corresponds
to small $x$ region.

In summary, the scaled factorial moments in second-order quark-hadron
phase transition are reexamined with the Ginzburg-Landau description, with
the spatial fluctuations of the phase angle of the field fully taken into
account. Scaling behaviors between $F_q$ and $F_2$ are shown, and the exponent
$\nu$ is found universal within accuracy $4\%$. It will be interesting to
investigate the effect of the full spatial fluctuations of the field on the
exponent $\nu$. This topic will be left for further discussions.
 
This work was supported in part by the NNSF, the SECF and Hubei NSF in China.

\newpage
\centerline{{\Large Figure Captions}}
\begin{description}
\item
{\bf Fig. 1} Dependences of $\ln F_q$ on the bin width $-\ln x$ after the
contribution from spatial fluctuations of the phase angle of the field fully 
taken into account (mode 2). Curves from lower to upper are for $q$ from 2 
to 8, respectively. 
\item
{\bf Fig. 2} Scaling behaviors of $\ln F_q$ vs
$\ln F_2$ for the same data as in Fig.1.
\item
{\bf Fig. 3} Upper part: $\ln F_q$ as functions of $-\ln x$ without spatial
fluctuations (mode 1); Lower part: Scaling behaviors between $\ln F_q$ and
$\ln F_2$, with the same data as in the upper part.

\item
{\bf Fig. 4} Scaling behaviors of $\ln \beta_q$ as function of $\ln (q-1)$
for the two modes.
\end{description}
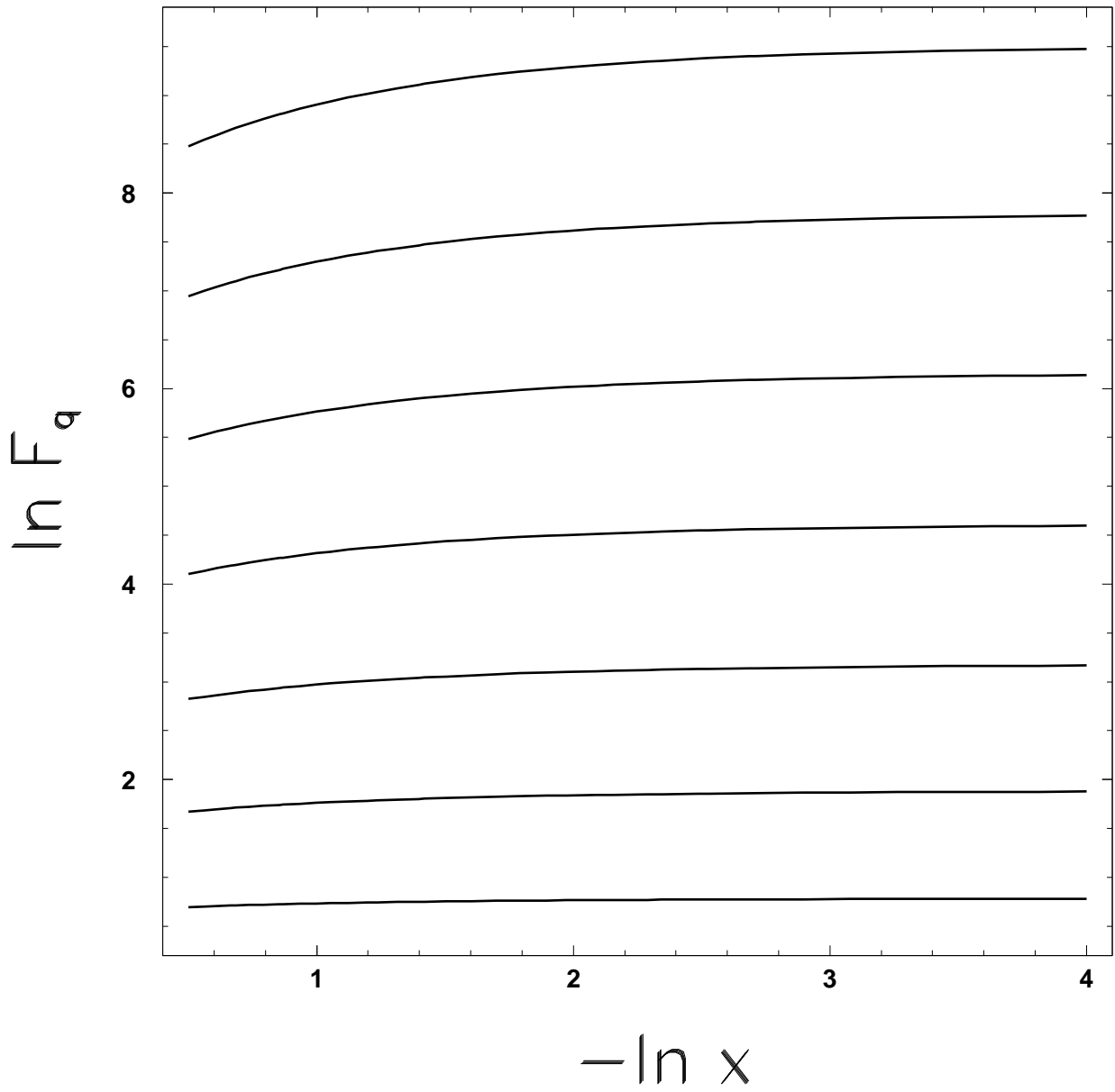
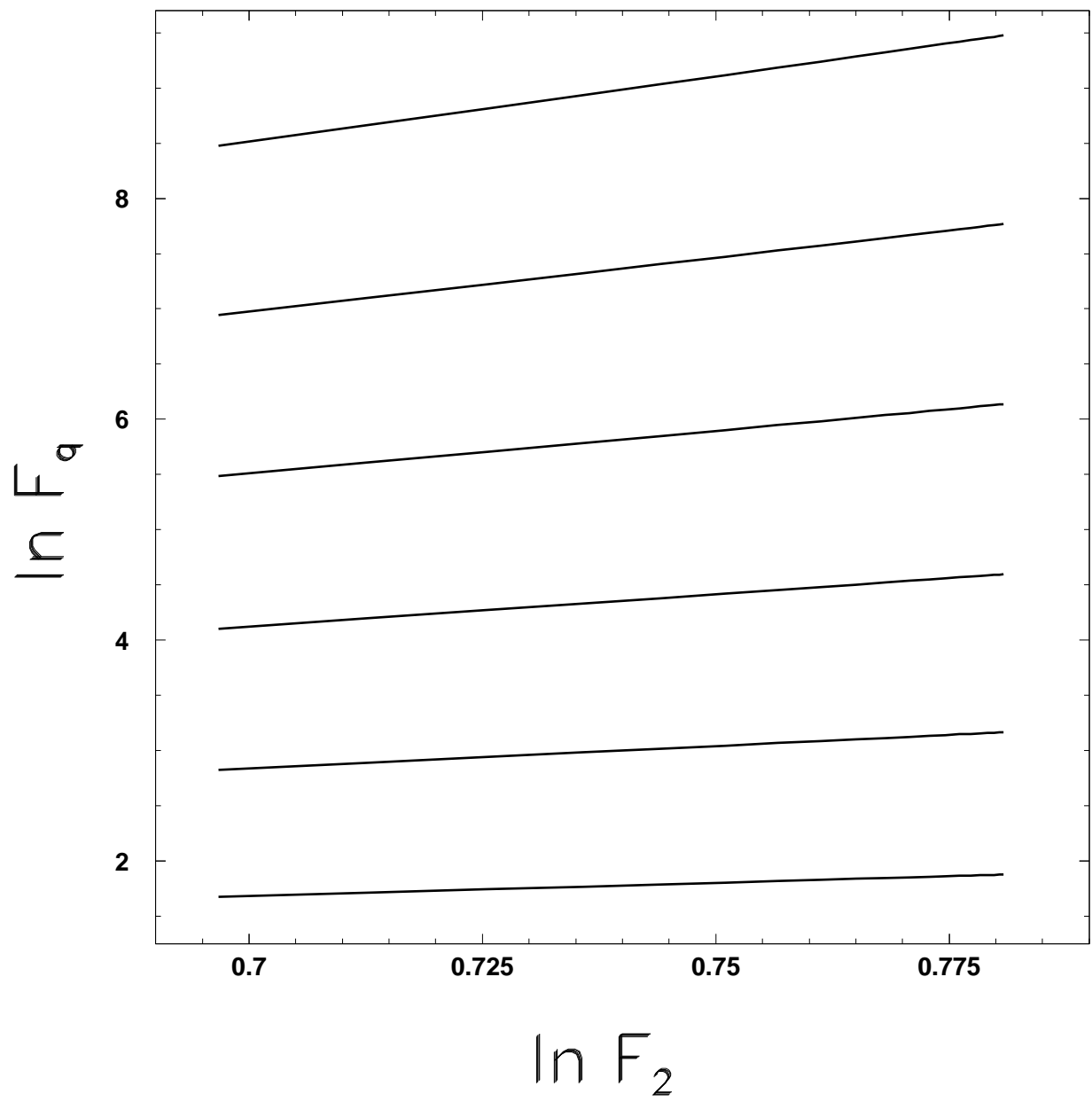
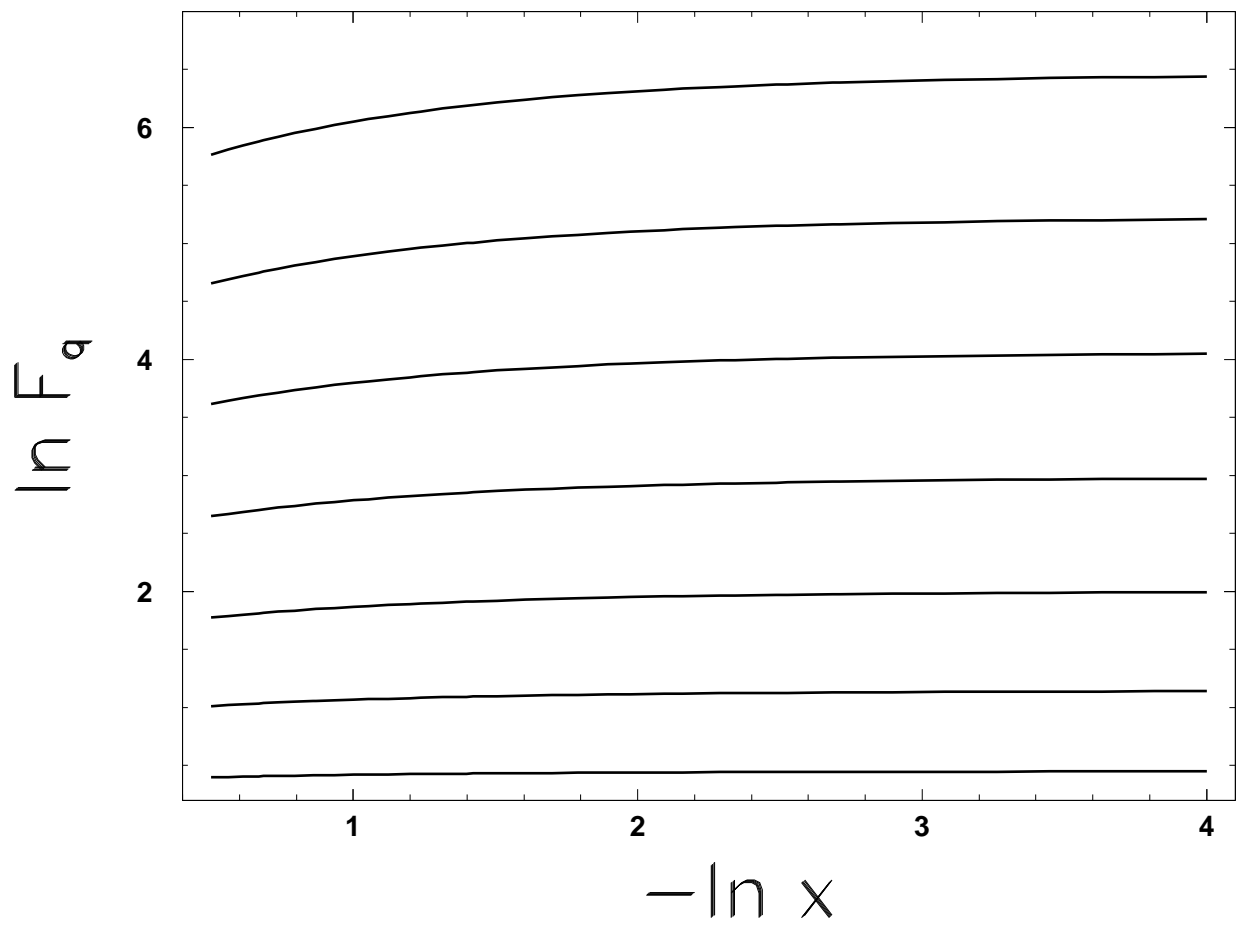
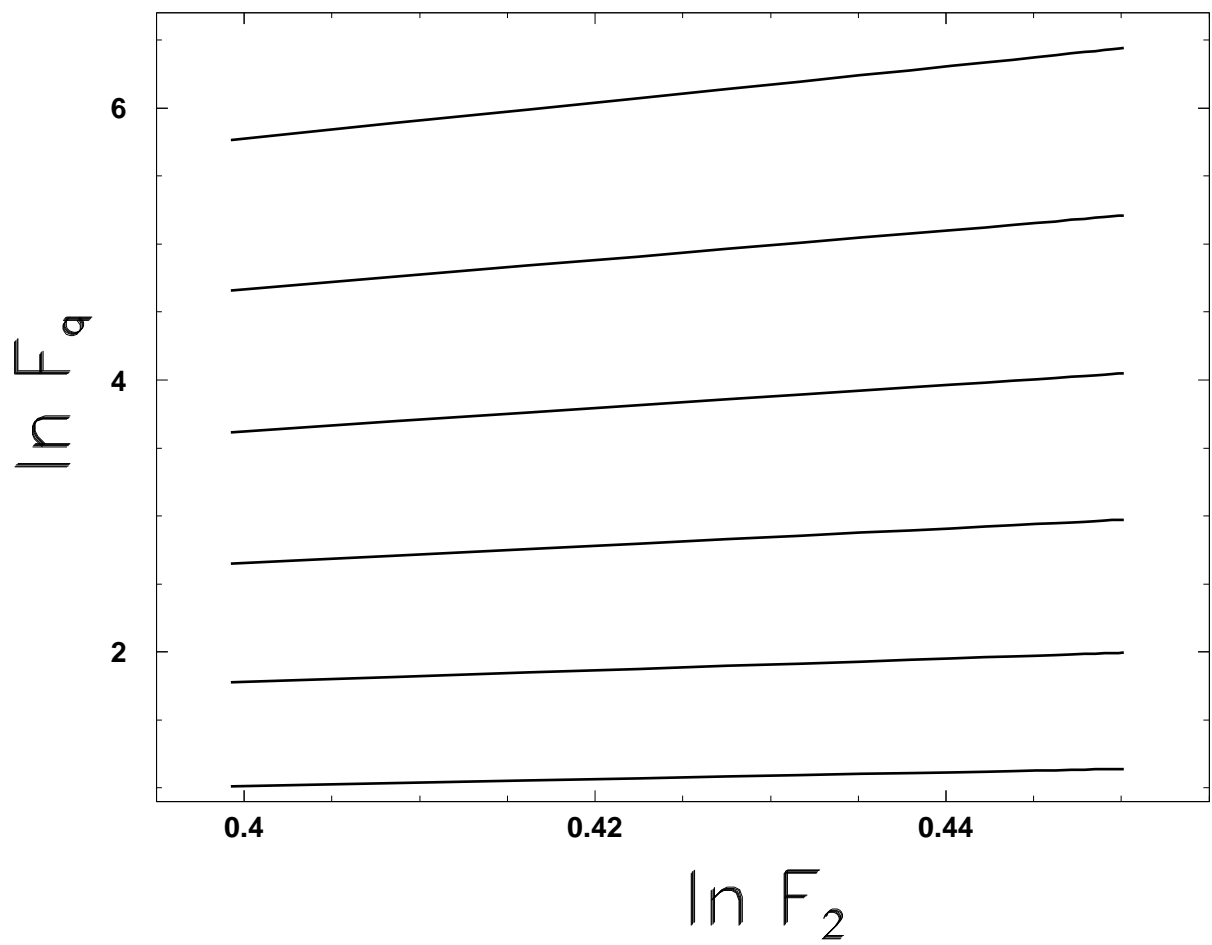
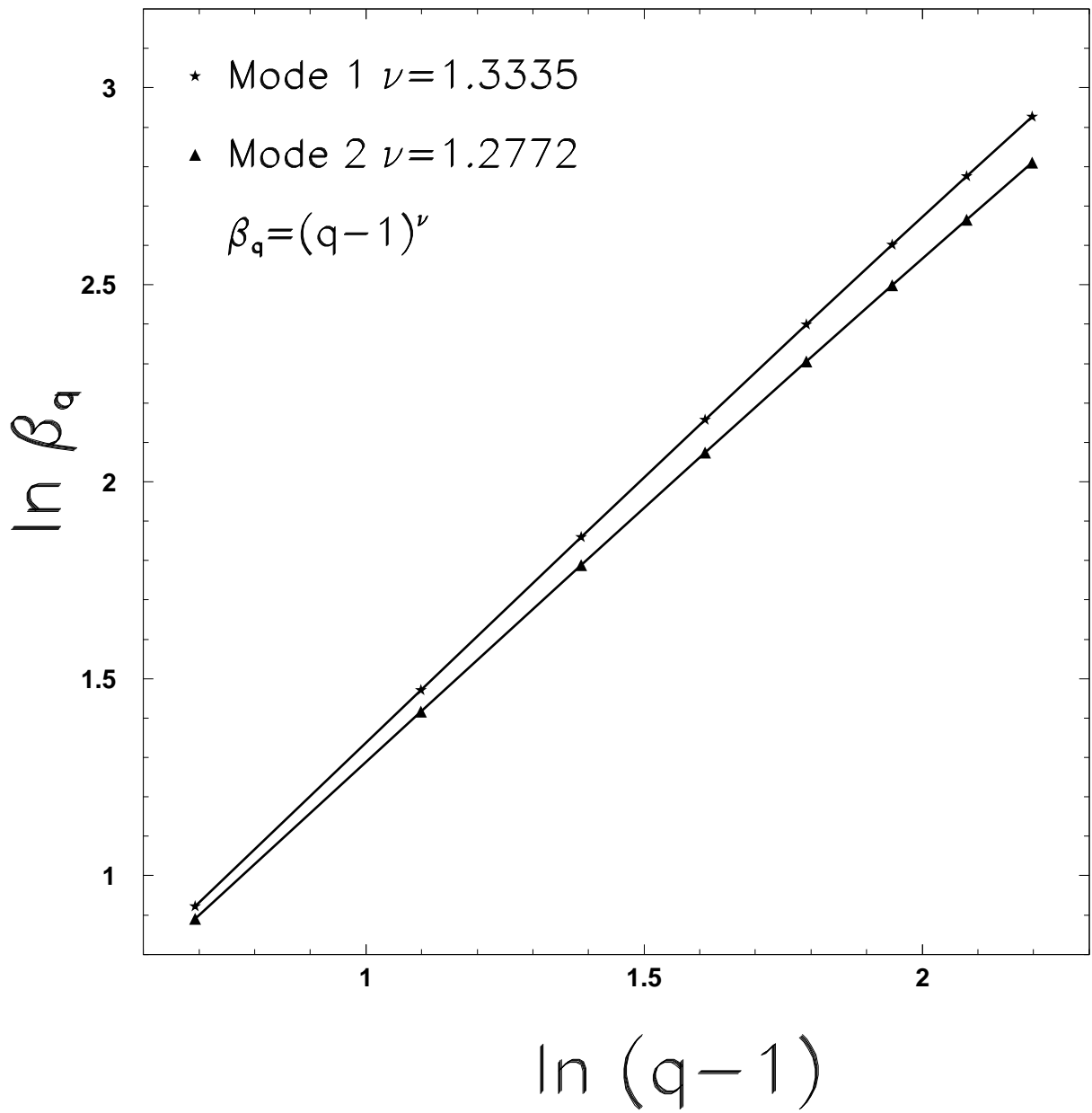
\end{document}